\documentclass[journal]{IEEEtran}

\usepackage{cite}
\usepackage{amsmath,amssymb,amsfonts}
\usepackage{algorithmic}
\usepackage{graphicx}
\usepackage{textcomp}
\usepackage{booktabs}
\usepackage[dvipsnames]{xcolor}
\usepackage[cachedir=.]{minted}
\usepackage{svg}
 
\def\BibTeX{{\rm B\kern-.05em{\sc i\kern-.025em b}\kern-.08em
    T\kern-.1667em\lower.7ex\hbox{E}\kern-.125emX}}

\usepackage{caption} %
\usepackage{subcaption} %

\newcommand{\edit}[1]{{\color{black}{#1}}}

\title{PowerPlots.jl:  An Open Source Power Grid Visualization and Data Analysis Framework for Academic Research}

\author{Noah~Rhodes,~\IEEEmembership{Member,~IEEE}%
\thanks{Support for this work was provided in part by the National Science Foundation (NSF) under the NSF CAREER award No. 2045860 and NSF ASCENT award No. 2132904, and by the Department of Energy through the LANL/LDRD Program and the Center for Nonlinear Studies.}%
\thanks{Noah Rhodes is with Los Alamos National Laboratory, Los Alamos, New Mexico, 87545 USA (email:{nrhodes@lanl.gov})}%
}

\date{August 2025}

\begin{document}

\maketitle

\begin{abstract}
    \edit{ Data visualization is essential for developing an understanding of a complex system. The power grid is one of the most complex systems in the world and effective power grid research visualization software must 1) be easy to use, 2) support unique data that may arise in research, and 3) be capable of creating custom figures for publication and presentation. However, no current software addresses all three of these needs. \emph{PowerPlots} is an open-source data visualization tool for power grids that does address these needs. In addition, several tools created to support this software facilitate the analysis of power grid data by transforming the data into graph topology or data-frame data formats that are more compatible for some analyses. In this work, we use \emph{PowerPlots} to investigate several case studies that involve exploring power grid data.  These case studies demonstrate the valuable insights that are possible when using network visualization and how it can be applied to research applications.}
\end{abstract}


\section{Introduction}

The power grid is a complex engineered network system and its operation is governed by the physics of power flow.  Developing an intuition for the operation of the power grid is a valuable skill that allows a researcher to identify when a unique behavior is occurring or more easily identify an error in their computational method.   Power grid visualization allows the researcher to see and recognize these patterns \cite{cuffe2015visualizing}.
\edit{Visualization of the power grid gives intuition through two use-cases. 1) \emph{Exploration}: it allows for exploration of data and debugging of methods by quickly and interactively visualizing data, and 2) \emph{Communication}: it enables sharing of specific information to communicate research findings.}

\edit{To support these use-cases, visualization software must be easy to use, support unique data that may arise in research, and be capable of creating custom figures for publication and presentation.}
\edit{Many packages, libraries, or ecosystems exist that can perform power grid visualization, but none have all three of these traits.  As a result, they may not be able to view and explore custom data, extend data formats with new information, and extensively customize figures for publication.}

\edit{Several existing open-source power grid modeling packages have a limited support for network visualization. They can show basic network structure and possibly show traditional power study data, but not unique research data from a novel power grid problem.}
\emph{PyPSA}~\cite{brown202516883512} or Python for Power System Analysis is an open source energy system modeling project.  It supports a  range of standard power grid optimization models like unit commitment and capacity expansion, and includes cross sector energy systems modeling.  It includes  plotting of networks for its users to visualize the results and view the solutions of these models.
\emph{OpenDSS}~\cite{opendss} is a distribution network simulator developed by EPRI.  Originally a scripting tool to define networks and routines for simulation, an extension for a graphical environment was developed, \emph{OpenDSS-G}, to enable a visual interface to create and view simulation results.  
\emph{PowerGridModel}~\cite{Xiang2023} is a an open source power grid analysis library that supports power flow, state estimation, and short circuit analysis. It includes visualization software in its data analysis extension~\cite{Schouten_PGM_DS}.  
\emph{PandaPower}~\cite{pandapower2018} is an open source python package for power system analysis.  It emphasizes a tabular data structure for data analysis of power flow, state estimation, and short-circuit calculations. It has basic support for visualizing the network representation of the data.

\edit{Many closed-source software ecosystems are even more limited in their ability to show unique research data. These tools typically have very limited flexibility to visualize data beyond what they are designed to calculate. }
\emph{PowerWorld}~\cite{powerworld} is an industry software initially developed to provide students with a visual tool to understand the power system concepts that they learned in classes.  The slogan ``the visual approach to electric power systems" emphasizes its roots as a tool to develop intuition outside of looking at charts and tables of power grid data.
A variety of other commercial industry software including PSSE~\cite{siemensKnowPSSE}, PLSF~\cite{gevernovaPowerFlow}, ETAP~\cite{etapGridModeling}, and TARA~\cite{powergemTARASoftware} have some level of visualization of network data like prices or  outage reliability.

These existing visualization software systems represent a mixture of open and closed source tools.  They provide visualization for projects within a specific ecosystem  and are generally for visualizing traditional well-studied power grid problems.  For users of these ecosystems who are analyzing traditional problems, these are excellent tools.  However, they are not always suitable or adaptable for research applications that solve novel power grid problems.

\emph{PowerPlots.jl} is a flexible tool for creating visualizations of power grids to better support the needs of researchers.  It has a simple interface for creating a network plot to view and interact with data, \edit{supports custom data fields,} and has a high level of customization to create publication-ready figures.
It is written in the Julia programming language~\cite{julia}, and adopts the data format from \emph{PowerModels.jl}~\cite{coffrin2018powermodels} and \emph{PowerModelsDistribution.jl}~\cite{fobes2020powermodelsdistribution}, which parse file formats including MatPower~\cite{zimmerman2010matpower}, PSSE~\cite{siemensKnowPSSE}, as well as a JSON dictionary.  The data format is extendable to visualize new components that are not natively supported by these file formats.  \edit{Furthermore,} several tools like a graph data model and a data-frame data model, created to support the network plots, are also useful as data analysis tools for research.  

Many grid research packages can take advantage of \emph{PowerPlots.jl} to enable analysis of their outputs with no change to their code \cite{rhodes2021powermodelsrestoration, powermodelsonm, Vanin2022, rhodes2020balancing}, and 
many publications already use \emph{PowerPlots.jl} to communicate their results in application areas including wildfire risk, carbon intensity metrics, restoration planning, and network reconfiguration~
\cite{gorka2025electricityemissions, yang2024multi, rhodes2023security, taylor2022framework, yang2024multistage, zhou2025machine, chen2024real, zhou2024mitigating, asiamah2025machine, asiamah2025developing}.

This paper is not documentation for \emph{PowerPlots.jl}; the documentation can be accessed at the GitHub repository of the package\footnote{https://wispo-pop.github.io/PowerPlots.jl/dev/}.
Rather, this paper explores how the package can be used as a research tool, including examples of the types of analyses that this package enables. 

The primary contributions of this paper are (1) An introduction to the plotting tool \emph{PowerPlots.jl} and its software design, (2) \edit{Several case studies that use} \emph{PowerPlots.jl} to explore network data and customize figures, and (3) An exploration of the types of data analysis that are provided by tools within \emph{PowerPlots.jl}.

The rest of the paper is organized as follows: Section \ref{sec:design} explains the design and basic software structure of \emph{PowerPlots.jl}.  Section \ref{sec:exploration} describes the features of \emph{PowerPlots.jl} to enable data exploration, with \edit{case studies} of how to investigate different types of research questions.  
Section  \ref{sec:analysis} shows how the data structures within \emph{PowerPlots.jl} can be utilized to conduct data analyses of power grid data.  Finally, Section \ref{sec:conclusion} concludes the work.

\section{Design of PowerPlots} \label{sec:design}

The two goals of \emph{PowerPlots.jl} are to provide easy exploration of power grid data, and to create clear communication of research results for publication and presentation.  To achieve these goals, 
the design of \emph{PowerPlots.jl} revolves around these principles: 1) simplicity, 2) flexibility, and 3) customization.  

The primary user of \emph{PowerPlots.jl} is a researcher exploring novel research problems, and these principles enable them to conduct their research more effectively.  
\emph{PowerPlots.jl} must be very simple to use with intuitive default behavior for initial data exploration about the network.  Flexibility in visualizations allows the user to visualize a wide range of custom information about a power grid. High levels of customization allows a user to make clear visualizations that highlight specific data for an audience.  

This section introduces how the software package's functions to support these principles. 
First, we introduce an overview of the package functionality. Next, we introduce the data models used by \emph{PowerPlots.jl} that also have utility for other analytical applications. 

\subsection{Plotting Process}
\emph{PowerPlots.jl} uses the nested dictionary structure of \emph{PowerModels.jl} as input.  The top level of the dictionary contains metadata about the power grid and a key for each of the power grid components.  Each grid component key refers to a dictionary containing data for each component.

\begin{minted}[breaklines,escapeinside=||,mathescape=false, linenos, numbersep=3pt, gobble=0, frame=lines, fontsize=\small, framesep=2mm]{julia}
grid_data = Dict(
"bus" => Dict(
  "1" => Dict(
    "vmax" => 1.1,
    "vmin" => 0.9,
    "base_kv" => 220,
    ...
  )
  "2" => Dict( ... )
    ...
"gen" => Dict( ... )
...
)
\end{minted}

The input data must be augmented with coordinate data for each component.
To generate coordinates, \emph{PowerPlots.jl}, creates a graph incorporating each of the component types in the data dictionary.  A graph layout algorithm is applied to this network to compute coordinate locations for each component.  Each component in the input data is updated with these coordinates.
Next, the input data must be converted from the nested dictionary structure to a set of data-frames.
The plotting engine used by PowerPlots is \emph{VegaLite.jl}~\cite{satyanarayan2016vega}, which takes a data-frame as input.  
Each component type (e.g. \mintinline{julia}{"bus"} or \mintinline{julia}{"gen"}) of the input data is converted into its own data-frame.  Each grid component is then plotted as a separate layer in a figure.
Finally, user arguments are applied to the figure.  These arguments might select which data field to use to color a component, change the size of the component, or add indicators to show direction of power flow. 

The default use of power plots automatically applies these steps to create a figure.  However, each step can be customized to enable control over the graph layout, the set of components included in the figure, and which data fields are used for the visualization.
If a user requires additional customization beyond what can be achieved with the user arguments, they can then modify the figure to fully customize the representation of e.g  legend placement, add data fields to visualize, or change the shape of grid components.  As a \emph{VegaLite.jl} figure, a wide variety of customization can be applied.

\subsection{Data Structures}
Using \emph{PowerPlots.jl} requires converting the nested dictionary data structure from \emph{PowerModels.jl} into a graph structure \mintinline{julia}{PowerModelsGraph} for computing network layouts and a DataFrame style structure \mintinline{julia}{PowerModelsDataFrame} for the \emph{VegaLite.jl} plotting backend. 

\subsubsection{PowerModelsGraph}
The graph structure created for \emph{PowerPlots} is an undirected graph of all of the edges and vertices seen in a visualization.  By default, network components including buses, generators, loads, and shunts are all nodes in the graph structure, and branches, dc lines, switches, and transformers are included as edges.  Nodes that are not buses are considered 'connected components' because they are directly connected to a bus, and an additional 'connector' edge type is added to represent these connection points. 

The data structure contains four fields: the undirected graph, as well as three mappings from the graph indices to the original power grid data.
\begin{minted}[breaklines,escapeinside=||,mathescape=false, linenos, numbersep=3pt, gobble=0, frame=lines, fontsize=\small, framesep=2mm]{julia}
mutable struct PowerModelsGraph
    graph::Graphs.SimpleDiGraph
    node_comp_map::Dict{Int,Tuple{String, String}}
    edge_comp_map::Dict{Graphs.AbstractEdge, Tuple{String, String}}
    edge_connector_map::Dict{ Graphs.AbstractEdge, Tuple{String, String}}
\end{minted}

A user can specify the components to include in the graph. This is useful for computing a network analysis without including information about generator connections.
The following example shows how to identify the highest node degree bus in a network by creating a graph where only buses are included as nodes.
\begin{minted}[breaklines,escapeinside=||,mathescape=false, linenos, numbersep=3pt, gobble=0, frame=lines, fontsize=\small, framesep=2mm]{julia}
using PowerPlots, PGLib, Graphs
grid_data = pglib("case1354")

pmg = PowerModelsGraph(grid_data; node_components=[:bus])
(v, id) = findmax(degree(pmg.graph)) # (14, 2)
pmg.node_comp_map[id] # (:bus, "1001")    
# the highest node degree is 14 at bus 1001
\end{minted}

\subsubsection{PowerModelsDataFrame}
The data structure used in PowerModels has a nested dictionary structure, with component types at the top level, followed by component indices,  with component parameters in a third dictionary. 
This structure is convenient for many of the common data look-ups and manipulations that are used when exploring power grid data associated with a single component.  However, it is not suitable to exploring aggregate data across a component type.

\emph{PowerPlots.jl} has a data structure to create and store a DataFrame of each component, the \mintinline{julia}{PowerModelsDataFrame}.  Metadata from the top level of the \emph{PowerModels.jl} dictionary is converted into its own Dataframe, while a DataFrame is created for each grid component.  Each of the grid components are stored in a dictionary for easy access.
\begin{minted}[breaklines,escapeinside=||,mathescape=false, linenos, numbersep=3pt, gobble=0, frame=lines, fontsize=\small, framesep=2mm]{julia}
mutable struct PowerModelsDataFrame
    metadata::DataFrames.DataFrame
    components::Dict{Symbol, DataFrames.DataFrame}
\end{minted}

This data-frame structure allows for easy querying of metrics about the components in the network. 
The component dictionaries can be accessed and viewed as follows.

\begin{minted}[breaklines,escapeinside=||,mathescape=false, linenos, numbersep=3pt, gobble=0, frame=topline, fontsize=\small, framesep=2mm]{julia}
using PowerPlots, PGLib
grid_data = pglib("case1354")
pmd = PowerModelsDataFrame(grid_data);
pmd.components[:gen]
\end{minted}
\begin{minted}[breaklines,escapeinside=||,mathescape=false, numbersep=3pt, gobble=0, frame=bottomline, fontsize=\small, framesep=2mm]{julia}
260×18 DataFrame
Row  pg    qg    gen_bus  pmax  ... 
1    6.66  1.33    124    10.0  ... 
2    6.66  1.35   2035    10.0  ... 
3    0.6   0.17   3390     1.2  ... 
4    0.4   0.12   1604     0.8  ... 
5    13.3  3.08   2446    20.0  ... 
6    0.8   0.22   5664     1.6  ... 
7    13.3  3.07   5481    20.0  ... 
...
\end{minted}

\section{Exploration of Grid Data} \label{sec:exploration}
The following section demonstrates several features of \emph{PowerPlots.jl} that enable exploration of grid data.  Each of the examples uses a network from PGLib~\cite{pglib} unless otherwise specified.  

\begin{figure}[t]
    \centering
    \includegraphics[width=0.7\linewidth]{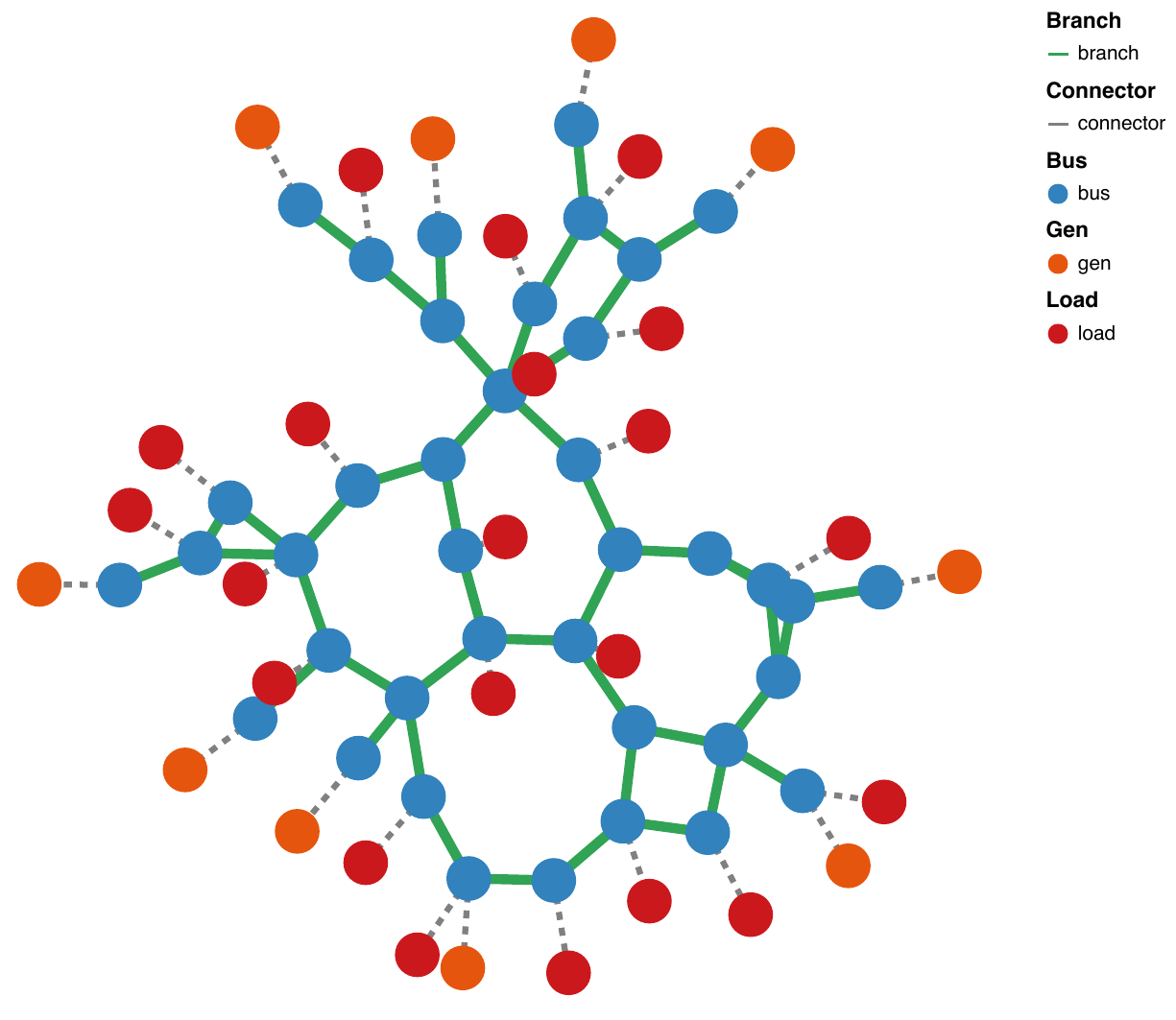}
    \caption{\small 39 Node EPRI Network}
    \label{fig:39}
\end{figure}

\subsection{Simplicity}
First, we demonstrate the simplicity of creating a power grid visualization. The following code example produces the visualization of the EPRI 39 node network in Fig. \ref{fig:39}.

\begin{minted}[breaklines,escapeinside=||,mathescape=false, linenos, numbersep=3pt, gobble=0, frame=lines, fontsize=\small, framesep=2mm]{julia}
using PowerPlots, PGLib
grid_data = pglib("case39_epri")
powerplot(grid_data)    
\end{minted}

This automatically includes the generator, load, branch, and bus components in the network and creates a network layout for visualization.  The default settings provide a view of the structure of the network, without visualizing any specific data about the components.

\begin{figure}[t]
    \centering
    \begin{subfigure}{1.0\linewidth}
        \centering
        \includegraphics[width=1.0\linewidth]{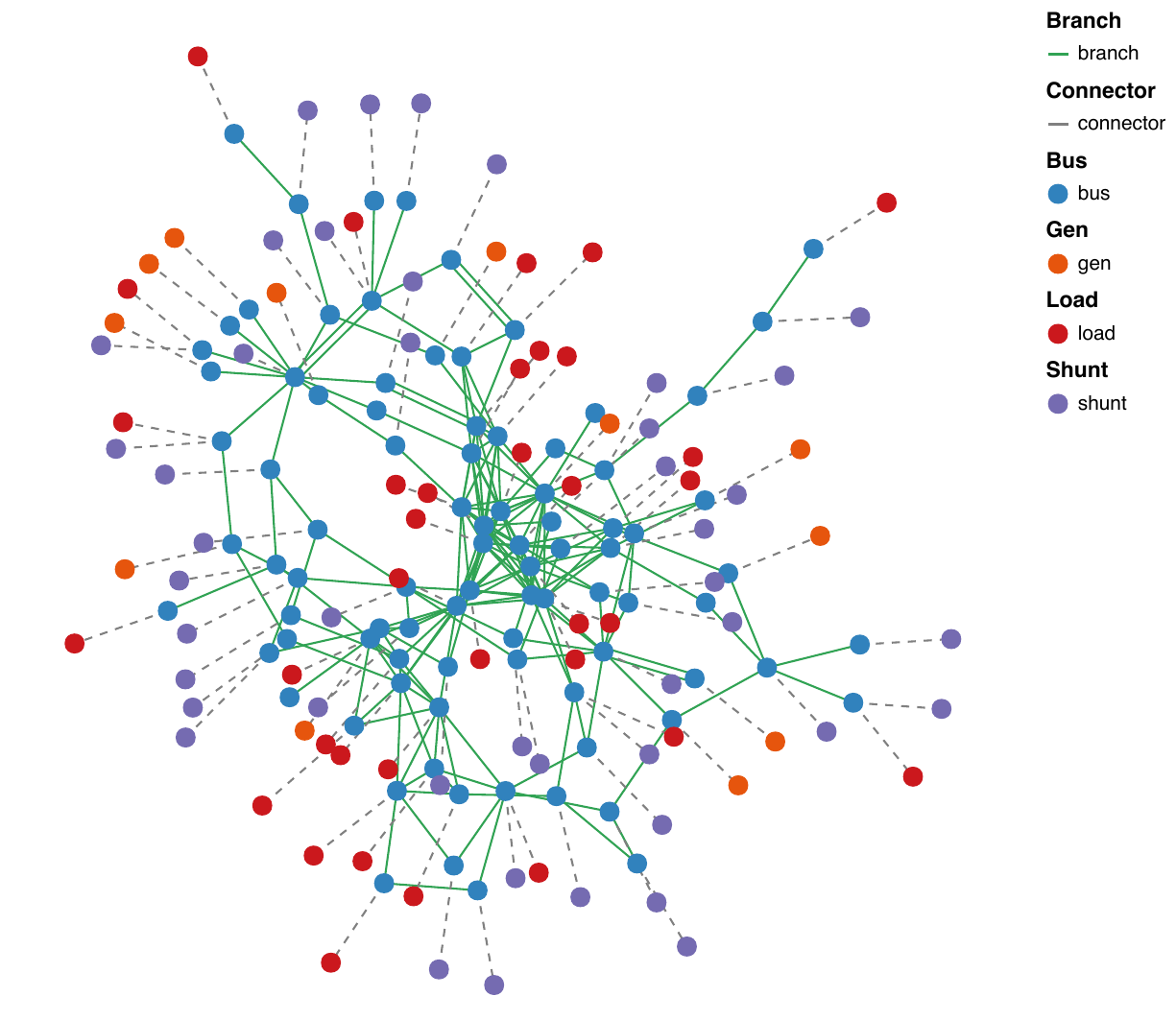}
        \caption{ \small 89 Node Pegase Network} \label{fig:89}
    \end{subfigure}
    \begin{subfigure}{1.0\linewidth}
        \centering
        \includegraphics[width=1.0\linewidth]{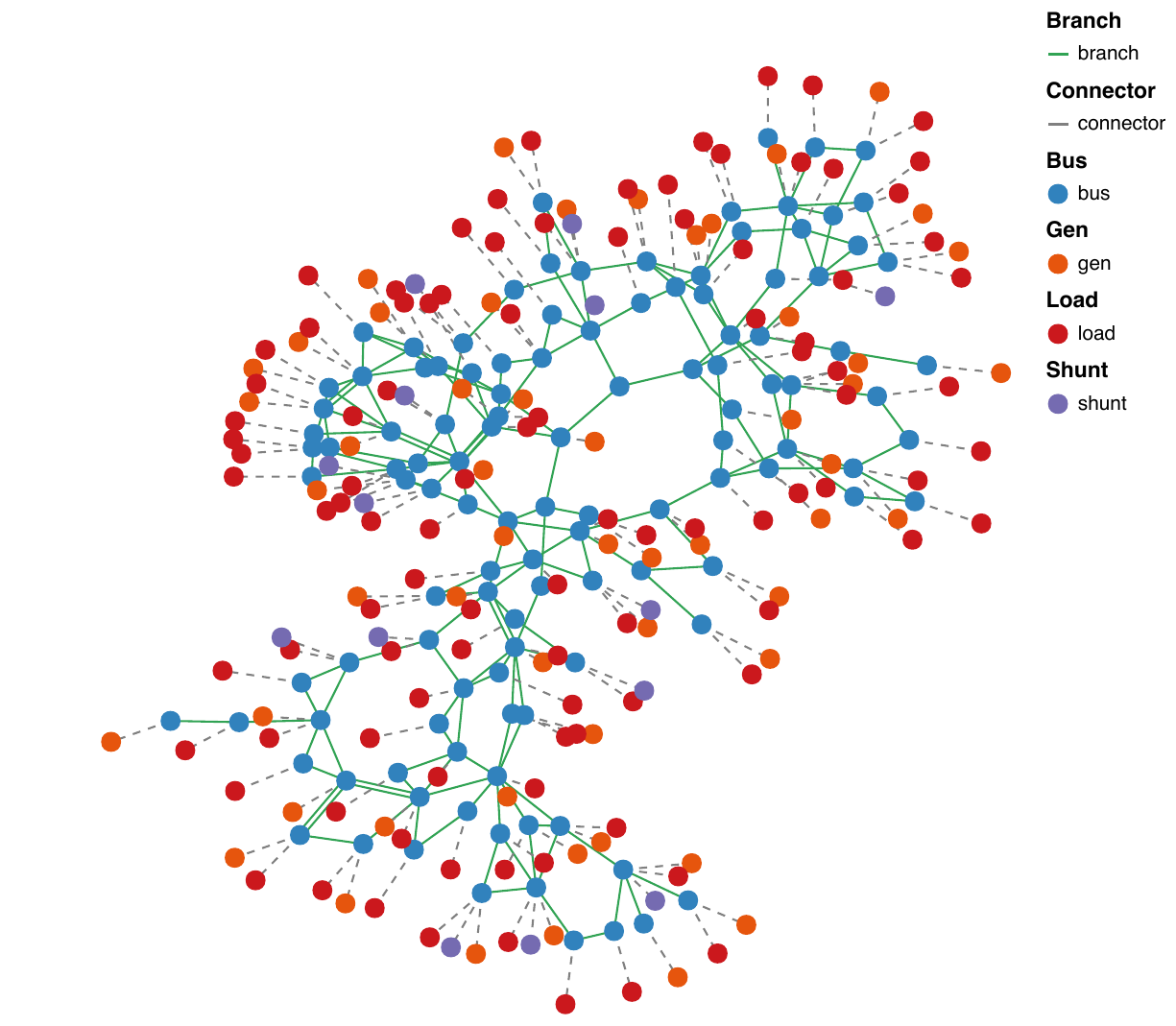}
        \caption{ \small 118 Node IEEE Network} \label{fig:118}
    \end{subfigure}
    \caption{\small It is quickly apparent that the 89 node network in Fig. \ref{fig:89} has a tight clustering of nodes in the center of the graph that are densely connected.  The 118 node IEEE network is shown in Fig. \ref{fig:118} as a comparison of a more typical network that lacks this density.} \label{fig:kron}
\end{figure}

Two larger networks are shown in Fig. \ref{fig:kron}, where the differences in the network structure are visually apparent.
In Fig. \ref{fig:89}, there is a very high-density collection of nodes in the center of the graph with a high degree of connectivity between them.  
In comparison, Fig. \ref{fig:118} shows a more typical transmission power grid network that does not have a high degree of connectivity in a subset of the network.  
This visual observation matches with the network analysis from \cite{kersulis2018topological}.  In this work, the authors analyze metrics like mean nodal degree, clique size, and adjacency spectra radius, and they estimate that some network processing such as Kron Reduction was applied to the PEGASE 89 node network along with several other PEGASE networks.  The same conclusion is evident from viewing the network topology.

Both of these networks are visually very busy with all of the generators, shunts, and loads included in the visualization, and a figure that is used in a paper should be more selective to highlight the network clustering.  However, this example demonstrates the default view when running the \mintinline{julia}{powerplot(grid_data)} command without additional arguments.

\begin{figure*}
    \centering
    \begin{subfigure}{0.24\textwidth}
        \centering
        \includegraphics[width=1.0\textwidth]{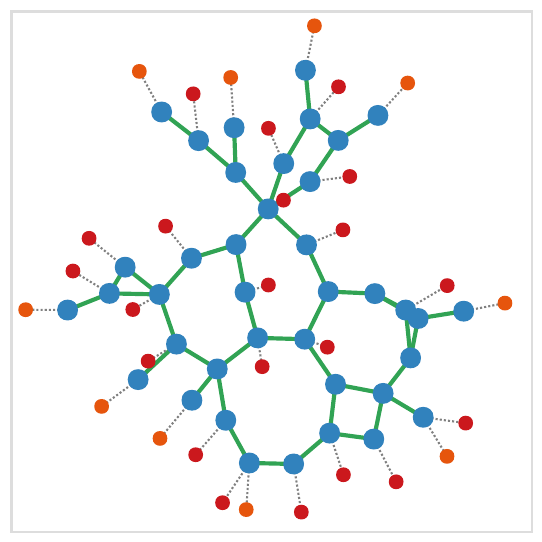}
        \caption{\small Kamada Kawai} \label{fig:kamada_kawai}
    \end{subfigure}
    \begin{subfigure}{0.24\textwidth}
        \centering
    \includegraphics[width=1.0\textwidth]{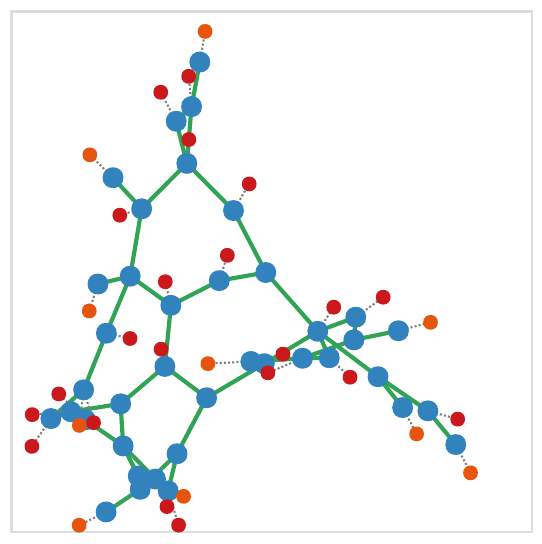 }
        \caption{\small Spring} \label{fig:spring}
    \end{subfigure}
    \begin{subfigure}{0.24\textwidth}
        \centering
        \includegraphics[width=1.0\textwidth]{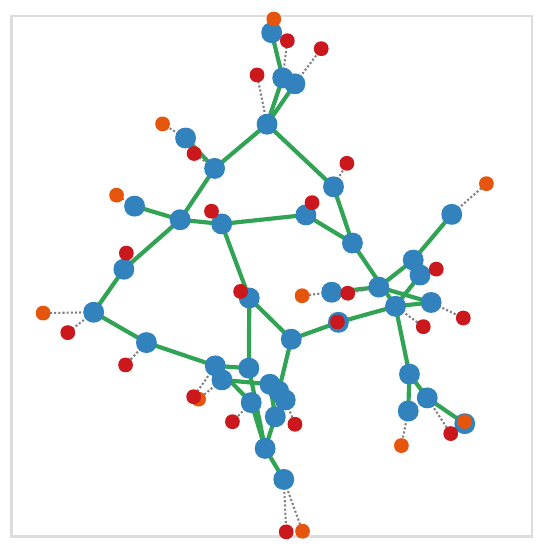}
        \caption{\small SFDP} \label{fig:SFDP}
    \end{subfigure}
    \begin{subfigure}{0.24\textwidth}
        \centering
        \includegraphics[width=1.0\textwidth]{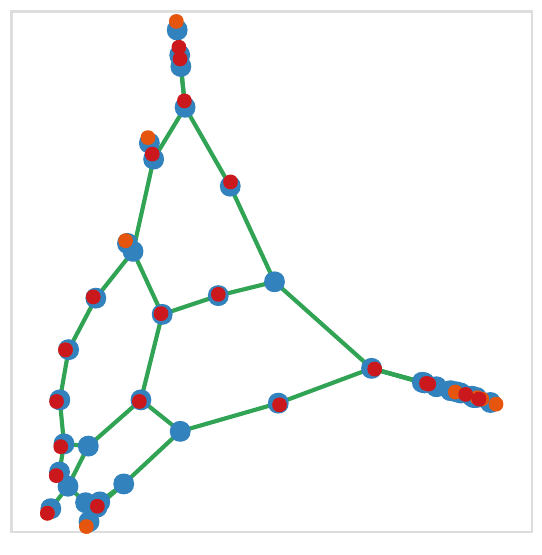}
        \caption{\small Spectral} \label{fig:spectral}
    \end{subfigure}
    
    \caption{\small Layouts of a 39 node network, sorted by time to compute the layout. The Kamada Kawai layout method generally produces the best visual layout to view the components of the network, but has a much longer computation time.  Buses are shown in blue, generators in orange, and loads in red.  Lines are shown in green, and dashed gray lines indicate where generators and loads connect to buses.}
    \label{fig:layouts}
\end{figure*}

\subsection{Layouts}

Many synthetic power grid networks used in research do not have labeled coordinates for the components of the power grid.  When the coordinates are missing, a network layout must be computed to visualize the power grid.

\emph{PowerPlots.jl} allows several types of network layout algorithms to be used, with varying tradeoffs of visual quality and computational speed.
Fig. \ref{fig:layouts} shows the layout algorithms available in \emph{PowerPlots.jl}.  
The default is the Kamada-Kawai algorithm which solves for the geometric distance between nodes  by minimizing the deviation from the graph distance between each node pair in the graph~\cite{kamada1989algorithm}.  
This  method produces layouts where each component of the power grid is clearly visible and typically has few edge crossings, however, it is the most computationally expensive of the available layout methods.  
The other layout methods are provided by \emph{NetworkLayout.jl}~\cite{networklayout2025}.  These methods include the Spring~\cite{fruchterman1991graph}, Scalable Force Directed Placement (SFDP)~\cite{hu2005efficient}, Spectral~\cite{koren2003spectral}, Shell, and Grid layouts.  Each of these methods has parameters, such as a spring force or edge weights, that can modify the resulting layout as seen below.

\begin{minted}[breaklines,escapeinside=||,mathescape=false, linenos, numbersep=3pt, gobble=0, frame=lines, fontsize=\small, framesep=2mm]{julia}
using PowerPlots, PGLib
grid_data = pglib("case39_epri")
powerplot(case; layout_algorithm=Spring, iterations=50)  
\end{minted}

\begin{table}[t]
    \centering
    \caption{Time to compute layout [sec]} \label{tab:layout}
    \resizebox{\linewidth}{!}{
        \begin{tabular}{r|lllll}
            \toprule\toprule
            Layout & Case39  & Case118 & Case500  & Case1354 & Case1888 \\ \midrule
            Kamada Kawai & 0.0096& 0.12& 2.3& 31& 35 \\
            Spring & 0.0039& 0.04& 0.48& 5& 4.8 \\
            SFDP & 0.00094& 0.006& 0.18& 3.1& 2.8 \\
            Spectral & 0.0012& 0.0085& 0.097& 2.5& 2.2 \\
            Shell & 0.00062& 0.0042& 0.013& 0.068& 0.065 \\
            Grid & 0.00055& 0.0022& 0.015& 0.25& 0.065 \\
            \bottomrule
        \end{tabular}
    }
\end{table}

Table \ref{tab:layout} shows the computational time for each layout method on networks from 39 to 1,888 buses.  On the largest network, the Kamada-Kawai layout requires 35 seconds to compute, while other methods may only require 2 to 4 seconds.  The Shell and Grid layouts are trivial to calculate and require much less than one second to compute. 

\begin{figure}[t]
    \centering
    \includegraphics[width=0.8\linewidth]{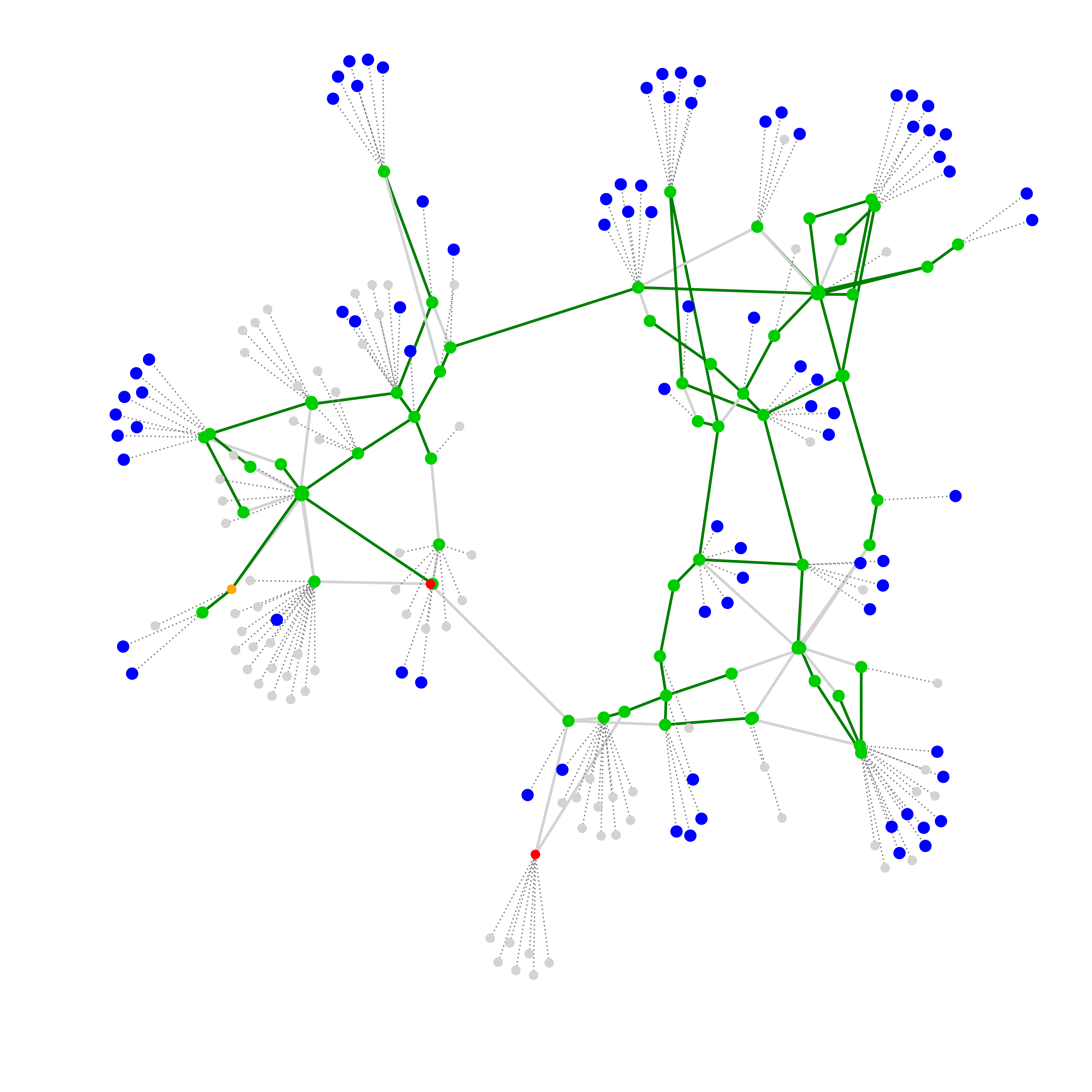}
    \caption{\small Partial fixed layout of power network.  Bus data is known, but generator locations are not.  The generator locations are computed using SFDP.  The figure shows a plan for a Public Safety Power Shutoff plan to reduce wildfire risk by turning off power lines. Components in gray indicate that they have been de-energized~\cite{rhodes2020balancing}. }
    \label{fig:fixed_layout}
\end{figure}

A layout is not required if node location data is available for the network.  If partial location data is available,  then the the SFDP layout can compute locations for the unknown nodes.  For example, Fig. \ref{fig:fixed_layout} shows the RTS-GMLC system~\cite{barrows2019ieee} which has location data for each bus, but the layout of generators is computed.
\begin{minted}[breaklines,escapeinside=||,mathescape=false, linenos, numbersep=3pt, gobble=0, frame=lines, fontsize=\small, framesep=2mm]{julia}
using PowerPlots, PGLib
grid_data = pglib("case39_epri")
# existing x and y coords are fixed
# missing coords are computed.
powerplot(grid_data; fixed=true)  
\end{minted}

A layout can also be precomputed and saved to the network data to avoid repeated computation for a large network. 

\begin{minted}[breaklines,escapeinside=||,mathescape=false, linenos, numbersep=3pt, gobble=0, frame=lines, fontsize=\small, framesep=2mm]{julia}
using PowerPlots, PGLib
grid_data = pglib("case39_epri")
grid_data = layout_network(grid_data; layout_algorithm=SFDP, C=0.1, K=0.9)
powerplot(grid_data; fixed=true)  
\end{minted}

\subsection{Interactivity}
\begin{figure}[t]
    \centering
    \begin{subfigure}{1.0\linewidth}
        \centering
        \includegraphics[width=0.7\linewidth]{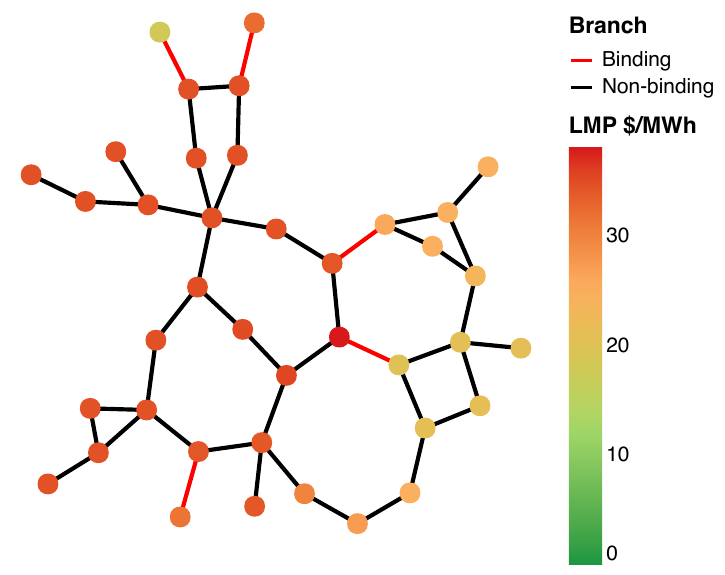}
        \caption{} \label{fig:lmp}
    \end{subfigure}
    \begin{subfigure}{1.0\linewidth}
        \centering
        \includegraphics[width=0.7\linewidth]{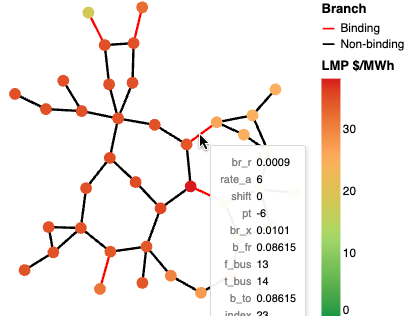}
        \caption{} \label{fig:lmp-interactive}
    \end{subfigure}
    \caption{\small Fig. \ref{fig:lmp} shows a plot of the Locational Marginal Price at each node, with binding transmission limits shown in red. In Fig. \ref{fig:lmp-interactive}, a user hovers over a binding transmission line to see detailed information.} \label{fig:lmp-main}
\end{figure}

A key value of \emph{PowerPlots.jl} in exploring a power grid data file is the ability to interactively view the grid information.  When hovering a mouse pointer over a power grid component in the plot, the associated information is shown, as seen in Fig. \ref{fig:lmp-interactive}.
Plotting a power grid can make it easy to visually identify patterns in the network, which can be further explored by hovering over the component to see valuable information.  

An example of the utility of this feature is shown in Fig. \ref{fig:lmp-main} where the locational marginal price (LMP) is plotted.  The LMP in a network can vary significantly as a result of congestion.  However, congestion is a difficult metric to gain intuition about solely from looking at numerical data.  Fig.  \ref{fig:lmp-main} shows the LMP for each node, and labels the congested branches in the network.   The congestion pattern is quickly recognizable and a user can then hover over a node to see the power demand of a  transmission line to find the parameter values of the line.  This provides useful information to develop a deeper analysis of the network congestion.

\subsection{Multi-Network Data}
Multi-Network data is a data representation in \emph{PowerModels.jl} for time series data, scenario data, multiple unique grid networks, or any other application where multiple representations of network data are utilized. 
The data file contains a new layer in the nested dictionary for each network in the multi-network data.  The following code block shows an example of the multi-network data:
\begin{minted}[breaklines,escapeinside=||,mathescape=false, linenos, numbersep=3pt, gobble=0, frame=lines, fontsize=\small, framesep=2mm]{julia}
multinetwork_data = Dict(
"nw" => Dict(
    "1" =>
        "bus" => Dict(
          "1" => Dict(
            "vmax" => 1.1,
            ...
          )
          "2" => Dict( ... )
            ...
        "gen" => Dict( ... )
        ...
      )
  "2" =>
        "bus" => Dict(
          "1" => Dict(
            "vmax" => 1.1,
            ...
          )
        ...
      )
\end{minted}

\begin{figure}
\centering
    \begin{subfigure}{1.0\linewidth}
        \centering
        \includegraphics[width=0.7\textwidth]{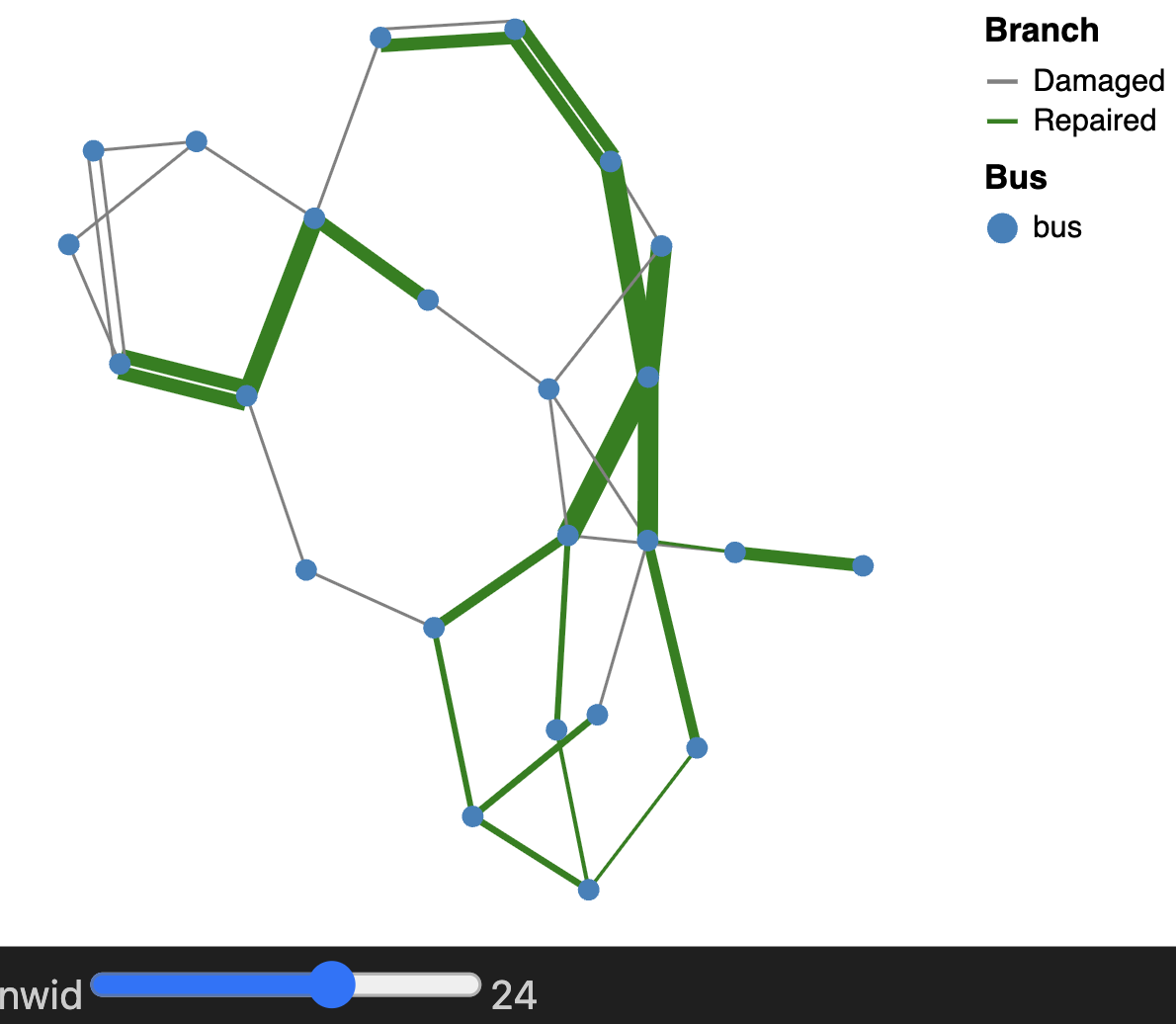}
        \caption{\small Repair stage 24} \label{fig:mn_1}
    \end{subfigure}
    \begin{subfigure}{1.0\linewidth}
        \centering
        \includegraphics[width=0.7\textwidth]{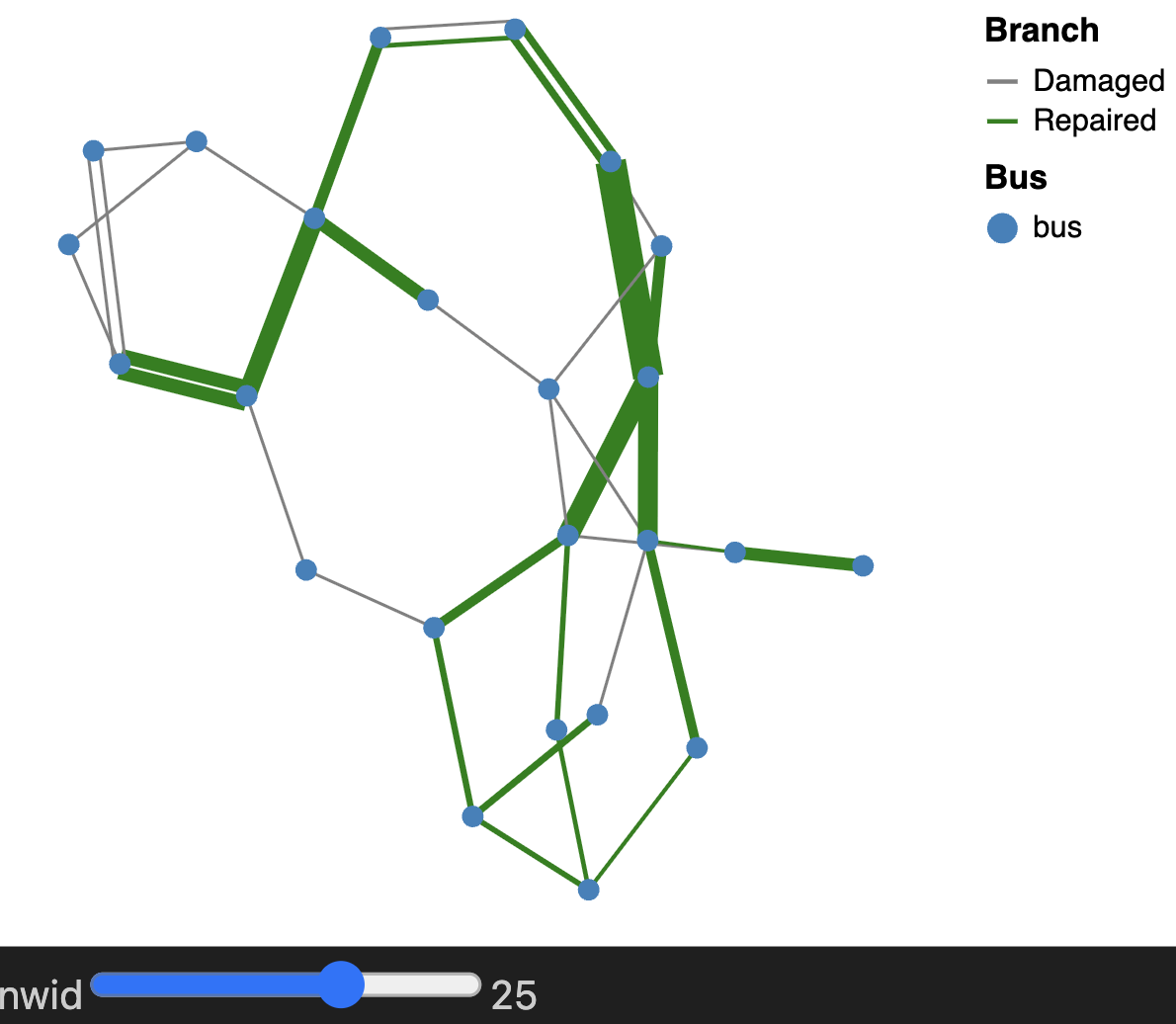}
        \caption{\small Repair stage 25} \label{fig:mn_2}
    \end{subfigure}
    \caption{\small Restoration plan applied to a 24-bus network.  Repaired branches are shown in green, while damaged lines are shown in gray.  The size of the branches indicates power flow.  The restoration plan created two islands which are connected to each other late in the repair process. The left island contributes significant power flow to the top center nodes after they are connected. }
    \label{fig:mn}
\end{figure}

\emph{PowerPlots.jl} uses this extra layer to create a filter in the plot to select which network is being viewed.  
This feature is useful, for example, in restoration planning to view the repair state of the network and the power flow at each stage of restoration.  
Fig. \ref{fig:mn} shows an optimized restoration plan implemented from~\cite{rhodes2021powermodelsrestoration} to optimize the order of repairs of damaged power lines to minimize total un-served energy.  The damaged branches in gray are sequentially repaired and turn green.  In addition, the amount of power flow  is indicated by the size of the branches. At repair step 24, two islands have been created, seen in Fig. \ref{fig:mn_1}.  Fig. \ref{fig:mn_2} shows repair 25, where a branch that connects these two islands is repaired and a significant amount of power flows from the left island to the right island.  
This indicates that the optimal solution for repair planning was to create multiple islands, before later connecting the islands.  

\subsection{Supported Networks and Components}
\begin{figure}[t]
    \centering
    \begin{subfigure}{1.0\linewidth}
        \centering
        \includegraphics[width=0.7\textwidth]{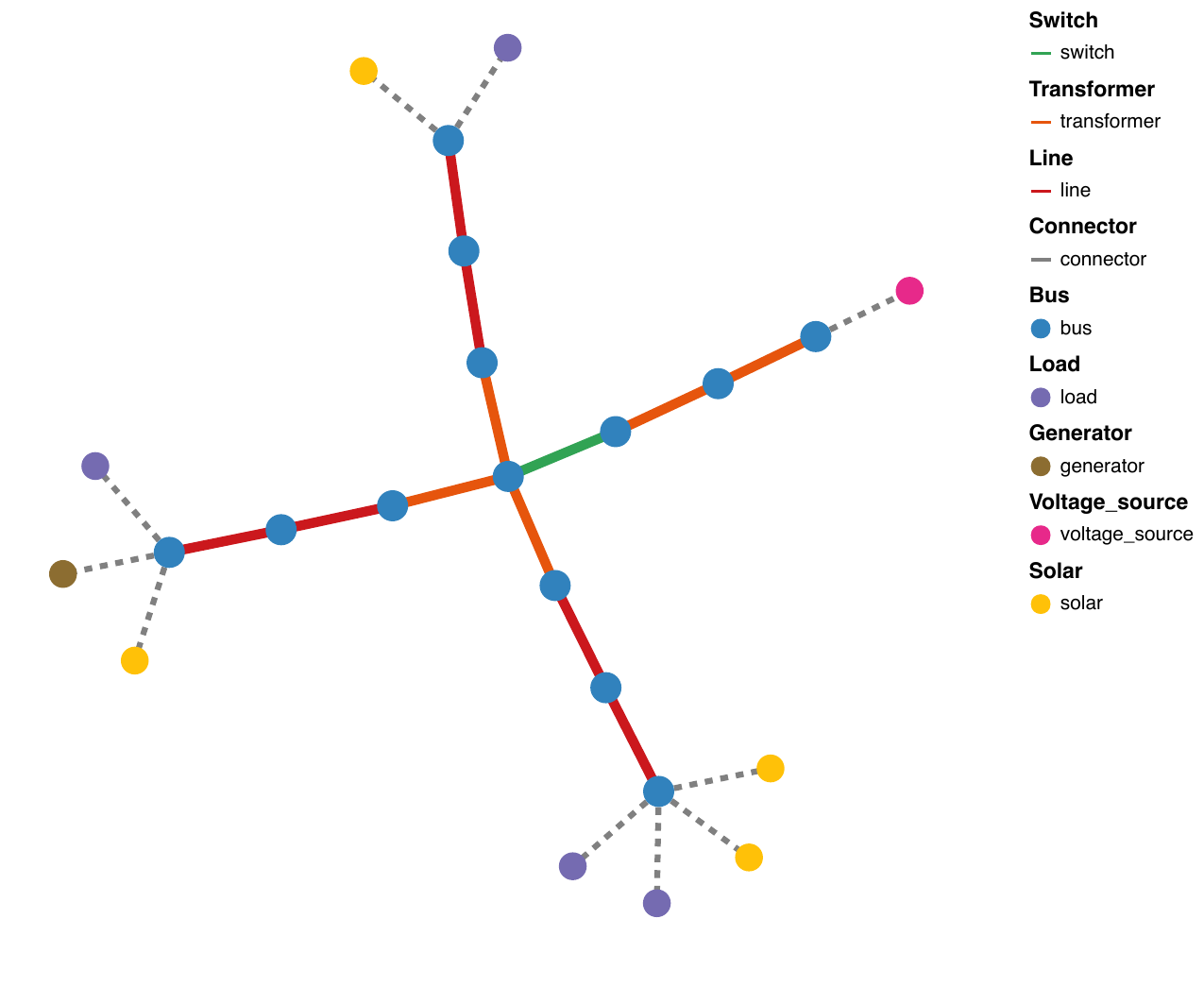}
        \caption{\small Engineering Model} \label{fig:eng}
    \end{subfigure}
        \begin{subfigure}{1.0\linewidth}
        \centering
        \includegraphics[width=0.7\textwidth]{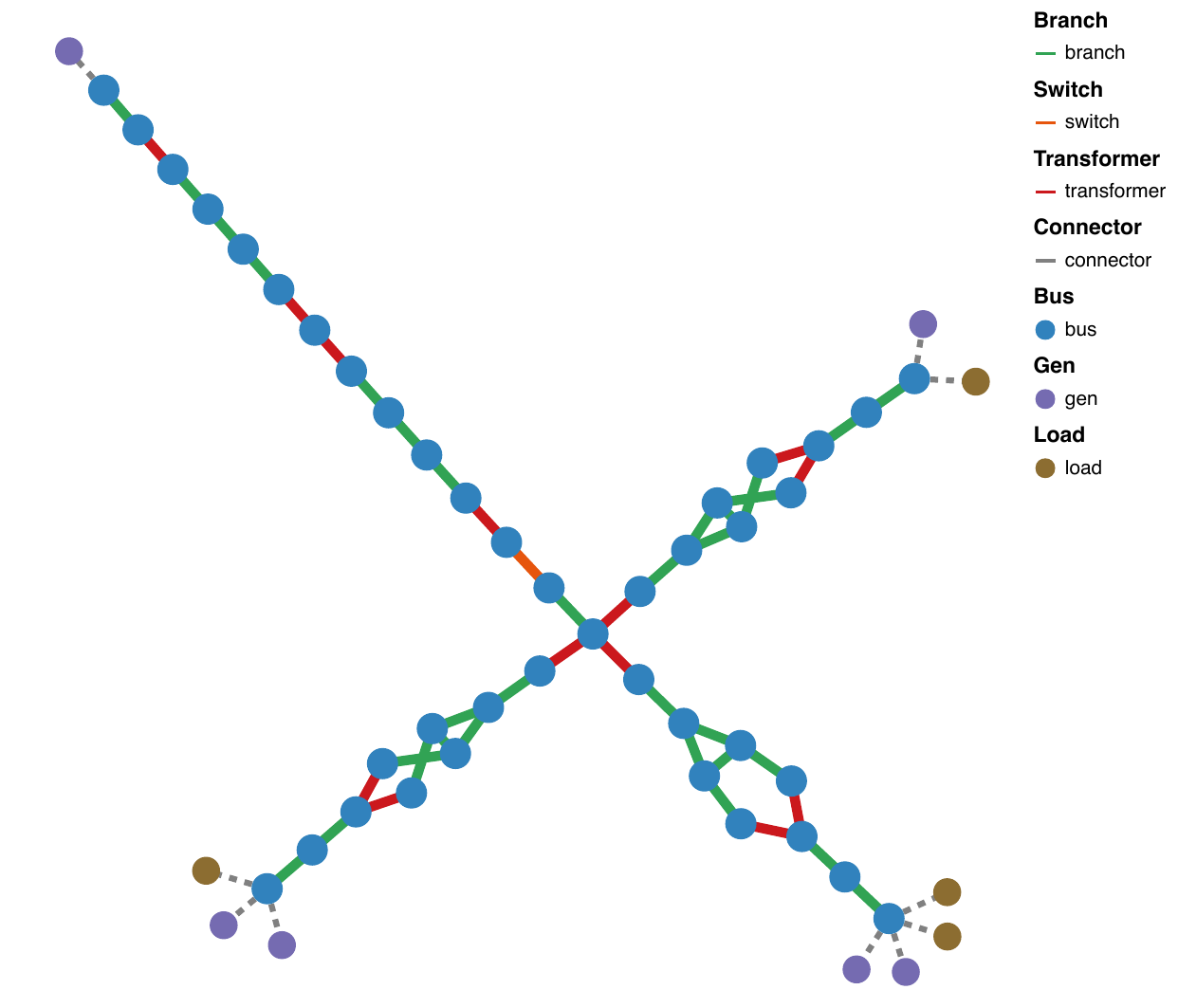}
        \caption{\small Math Model} \label{fig:math}
    \end{subfigure}
    \caption{\small Distribution grids have a engineering model and a mathematical model.  Shown here is the equivalent circuit of a 3-three center tap transformer.  }
    \label{fig:distribution}
\end{figure}

While originally designed for transmission networks, \emph{PowerPlots.jl} also supports \emph{PowerModelsDistribution.jl}~\cite{fobes2020powermodelsdistribution} for distribution network visualization.   \emph{PowerModelsDistribution.jl} has two data models, an engineering model representing the actual components of the power grid, and a mathematical model representing an equivalent network that is used for its optimization problems.  An example of the two network representations is shown in Fig. \ref{fig:distribution}. 

This figure also shows the large number of component types that can be visualized. By default, all of these components are included in the figure.  A subset of those components can be specified, and any additional components can also be specified. For example,  if a researcher is creating a problem with a new component in the data model, e.g "hydro", 
they can plot the new component with

\begin{minted}[breaklines,escapeinside=||,mathescape=false, linenos, numbersep=3pt, gobble=0, frame=lines, fontsize=\small, framesep=2mm]{julia}
powerplot(data, connected|\textunderscore|components = [:hydro, :gen, :load])
\end{minted}

\subsection{Customization} \label{sec:custom}
Many of the features of \emph{PowerPlots.jl} allow for quick visualization of a network to gain an understanding of the structure and patterns in the system.  When communicating the information to an audience, a user will often reduce the information shown in the plot to create a figure tailored to the topic of their research.

Because \emph{PowerPlots.jl} uses the plotting library \emph{VegaLite}, it is possible to take advantage of the many features in \emph{VegaLite} to customize a final figure.  
Here, we take the LMP figure from Fig. \ref{fig:lmp-main} and show the changes from a basic plot to a polished figure.

First, we create a plot with the following specification.  The code to compute the LMP is not shown for brevity.   As seen here, no connected components are shown, the size of the figure is set, and data fields of the bus and branch components are specified alongside the color scheme.  These modifications are enabled by passing user arguments to \emph{PowerPlots.jl}.

\begin{minted}[breaklines,escapeinside=||,mathescape=false, linenos, numbersep=3pt, gobble=0, frame=lines, fontsize=\small, framesep=2mm]{julia}
p = powerplot(data, connected_components=Symbol[],
    height=250, width=250,
    bus=(
     :size=>100,
     :data=>:LMP,
     :data_type=>:quantitative,
     :color=>reverse( colorscheme2array(
       ColorSchemes.colorschemes[:RdYlGn_4]
     )),
    ),
    branch=(
     :size=>2,
     :data=>:loading,
     :data_type=>:nominal,
     :color=>["red","black"]
    ),
)
\end{minted}

\begin{figure}[t]
    \centering
    \includegraphics[width=0.965\linewidth]{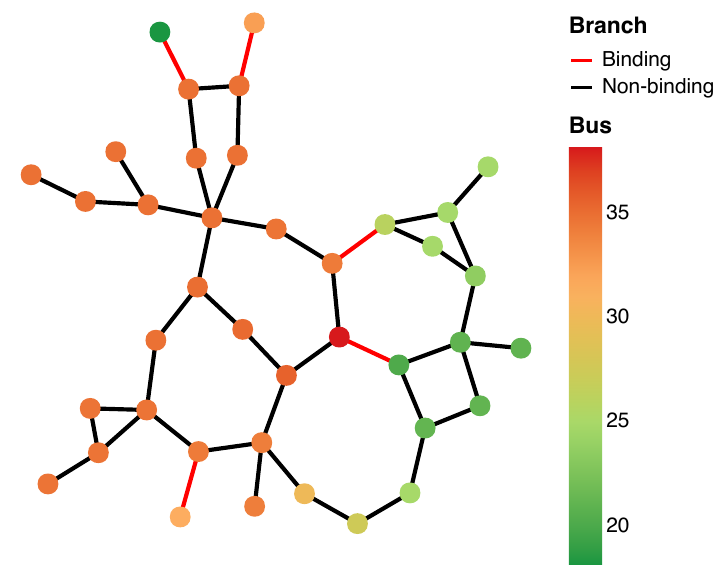}
    \caption{ \small LMP figure with no customizations made of the output of the `powerplot()` function .}
    \label{fig:intial}
\end{figure}

The figure created by this code is shown in Fig. \ref{fig:intial}.  It is very similar to the image from Fig. \ref{fig:lmp} with some exceptions: 1) The legend titles are the component names, instead of the data they show. 2) The domain of the bus color range starts from $\approx 18$.
These are changes that would need to be modified to accurately communicate results to a reader or viewer.  

The following code edits the plot to make these changes and create the figure as shown in Fig. \ref{fig:lmp}.
\begin{minted}[breaklines,escapeinside=||,mathescape=false, linenos, numbersep=3pt, gobble=0, frame=lines, fontsize=\small, framesep=2mm, breakafter=\] ]{julia}
p.layer[1]["layer"][1]["encoding"]["color"] ["legend"] = Dict("title"=>"% Loading")
p.layer[2]["encoding"]["color"]["scale"] ["domain"] = [0, max_LMP]
p.layer[2]["encoding"]["color"]["legend"] = Dict("title"=>"LMP \$/MWh")
\end{minted}

Additional modifications like modifying the size of each bus according to the power demand at the node, altering the height of the legend color gradient, and changing the font of the legend to match the manuscript is shown below, and the resulting figure shown in Fig. \ref{fig:final}.
\begin{minted}[breaklines,escapeinside=||,mathescape=false, linenos, numbersep=3pt, gobble=0, frame=lines, fontsize=\small, framesep=2mm, breakafter=\] ]{julia}
p.layer[2]["encoding"]["color"]["legend"] ["gradientLength"]=80
p.layer[2]["encoding"]["size"]  = Dict{String,Any}("field"=>"load", "type"=>"quantitative")
p.layer[2]["encoding"]["size"]["legend"] = Dict("title"=>"Load p.u.")
p.layer[2]["encoding"]["size"]["scale"] = Dict("range"=> [50,400])

using Setfield
@set! p.resolve.scale.size = :independent

font = "Times New Roman"
for x in ["labelFont", "titleFont"]
    p.layer[1]["layer"][1]["encoding"] ["color"]["legend"][x] = font
    p.layer[2]["encoding"]["color"] ["legend"][x] = font
    p.layer[2]["encoding"]["size"] ["legend"][x] = font
end
\end{minted}

\begin{figure}[t]
    \centering
    \includegraphics[width=1.0\linewidth]{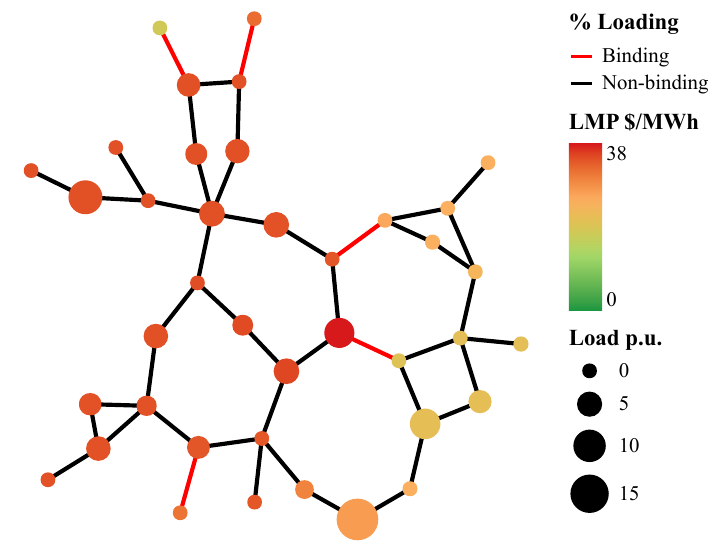}
    \caption{\small Power grid showing LMP by node color and nodal load by node size, with binding transmission limits in red. The legend titles have been modified, the limits of the color range for the LMP has been set to $[0,  \max\{LMP\}]$, and the height of the color range has been reduced.  The font in the legend has been changed to match the font of this manuscript.}
    \label{fig:final}
\end{figure}

\subsection{Summary}
The various features in this section highlight the many ways to use and interact with \emph{PowerPlots.jl}.  These features enable research by being easy to use for initial data exploration, allowing for flexibility in the data visualization, and supporting customized visualizations for publication.  

\section{Data Analysis} \label{sec:analysis}
The data structures in \emph{PowerPlots.jl} can also be used to enable data analysis of power grid data.  This can provide insights into the network structure or patterns of data that otherwise can be difficult to identify. 

\subsection{Graph Analysis}
Building a \mintinline{julia}{PowerModelsGraph} of the input data creates a graph structure of the network.  This graph structure can be used to get an incidence matrix of the network, compute shortest paths in the network, or other network computations from the \emph{Graphs.jl} package \cite{Graphs2021}.

\begin{figure}[t]
    \centering
    \includegraphics[width=0.95\linewidth]{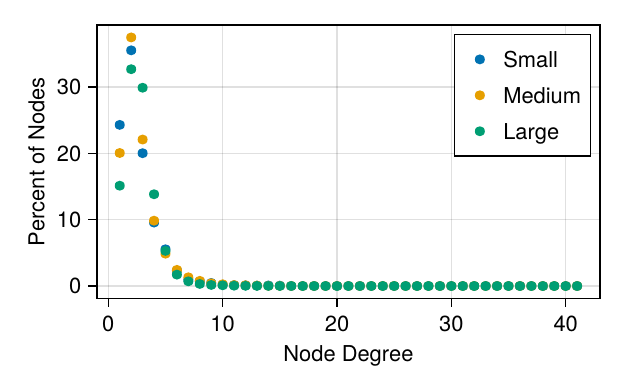}
    \caption{\small Distribution of bus node-degree across all networks in PGLib.  Small networks are less than 1,000 buses, medium between 1,000 and 10,000 buses, and large networks are greater than 10,000 buses.}
    \label{fig:node_degree}
\end{figure}

\emph{Example:} The distribution of node degrees in synthetic networks can vary with the system size.  Fig. \ref{fig:node_degree} shows the node degree $n$ for small ($n<1000$), medium ($1000<n<10,000$), and large ($n>10,000$) power grids from the PGLib~\cite{pglib} synthetic network library.  Small and medium sized networks have an increased proportion of one and two degree nodes, but fewer three and four degree nodes compared to the large networks.  The maximum node degree among the small networks is 15, while the medium and large networks both have a maximum node degree of 41.  The distribution of node degrees may have an impact on the results of some analyses, such as cascading effects in power grids.  Further investigation in the difference between these synthetic networks and the topology of real-world power grids may be useful for understanding where existing synthetic networks can provide meaningful results~\cite{cotilla2012comparing}.

\subsection{Tabular Data Analysis}
While the use case of the \mintinline{julia}{PowerModelsDataFrame} data structure in \emph{PowerPlots.jl} is to organize data for the plotting backend,  this format can be useful for analyzing statistics across components within the network.  After constructing a data-frame of the network data,  data-frame manipulations can be done using the \emph{DataFrames.jl} package~\cite{bouchet2023dataframes}. 

\emph{Example:} The following example finds the indices of the largest 5 generators in the network. 

\begin{minted}[breaklines,escapeinside=||,mathescape=false, linenos, numbersep=3pt, gobble=0, frame=lines, fontsize=\small, framesep=2mm]{julia}
pmd = PowerModelsDataFrame(data)
sort!(pmd.components[:gen].pmax)
pmd.components[:gen][1:5,:index]
# the five highest capacity generators are
# 1, 54, 101, 41, and 65 
\end{minted}

\emph{Example:} The following example explores metrics on the voltage magnitude at buses by voltage level.  The code example compute the mean, standard deviation, minimum, and maximum voltage for each group of voltage levels.

\begin{minted}[breaklines,escapeinside=||,mathescape=false, linenos, numbersep=3pt, gobble=0, frame=lines, fontsize=\small, framesep=2mm]{julia}
data = pglib("pglib_opf_case2000")
result = solve_ac_opf(data, Ipopt.Optimizer)
update_data!(data, result["solution"])
pmd = PowerModelsDataFrame(data)
gdf = groupby(pmd.components[:bus], :base_kv)
sort(combine(gdf, nrow, :vm => mean, :vm => std, :vm => minimum, :vm => maximum), :base_kv)
\end{minted}

Table \ref{tab:vm} shows these metrics. 
We find that the voltage is noticeable higher on the 69 kv buses compared to the high voltage network or to the low voltage network. This could  be indicative of a significant amount of power injection occurring at that voltage level in this network.

\begin{table}[t]
\centering
\begin{tabular}{cccccc} \\
Base kv & Bus Count & $|V|$ Mean & $|V|$ Std & $|V|$ Min & $|V|$ Max \\ \hline 
13& 4& 1.05& 0.0414& 1& 1.1 \\
14& 121& 1.04& 0.0441& 0.9& 1.1 \\
18& 33& 1.05& 0.0324& 0.933& 1.1 \\
20& 14& 1.04& 0.045& 0.935& 1.09 \\
22& 13& 1.04& 0.0482& 0.906& 1.09 \\
24& 7& 1.03& 0.0521& 0.934& 1.08 \\
69& 686& 1.07& 0.0126& 1.03& 1.1 \\
100& 418& 1.03& 0.0523& 0.911& 1.1 \\
138& 526& 1.06& 0.0145& 1.02& 1.09 \\
345& 178& 1.06& 0.0319& 0.929& 1.09 \\
\end{tabular}
\caption{Bus voltage magnitude metrics aggregated by bus voltage.}\label{tab:vm}
\end{table}

\section{Conclusion} \label{sec:conclusion}
Visualization of data is a powerful tool for building an understanding of data and highlighting aspects for communication.  
\emph{PowerPots.jl} has a simple interface to construct a view of the network information, but has significant flexibility to enable visualization of any parameter of interest.  It also allows significant customization to enable a user to create a figure that highlights the important information they wish to communicate. 

In creating \emph{PowerPlots.jl}, several tools are developed that also provide utility in data analysis of the power grid.  These include data transformations to graph network models and to tabular data formats that make applicable data analyses easier to perform.  
Together, the visualization and analysis tools in \emph{PowerPlots.jl} enable researchers to explore novel power grid problems by providing capabilities that support research beyond standard power grid data.

This software package will continue to evolve and develop new features.  Additional customization features will be available from the user arguments, and the implementation of these features will be guided by community involvement in this package. Furthermore, 
future extensions of this work can be applied to visualizing other network systems such as gas, water, or telecommunications networks.  This will eventually enable a visual understanding for cross-sector critical infrastructure analysis.  

\section*{\small Acknowledgments}
\small{
I would like to thank my PhD advisor Line Roald for her encouragement in developing this package, Hanbin Yang, Kshitij Girigoudar, and Joe Gorka for being early users of this package, and the people who have participated in the open source project by reporting issues and giving suggestions online.  I would also like to thank the Advanced Network Science Initiative at Los Alamos National Laboratory for creating the \emph{InfrastructureModels.jl} ecosystem on which \emph{PowerPlots.jl} was developed.
}

\bibliographystyle{IEEEtran}
\bibliography{IEEEabrv,ref}

LA-UR-25-29757

\end{document}